\pdfoutput=1

\documentclass[12pt,a4paper]{article}

\usepackage{ifthen} 
\usepackage{graphicx}  
\usepackage{epstopdf}
\newboolean{pdflatex}
\setboolean{pdflatex}{true} 

\newboolean{articletitles}
\setboolean{articletitles}{true} 

\newboolean{uprightparticles}
\setboolean{uprightparticles}{false} 

\usepackage{booktabs} 
\usepackage{microtype}
\usepackage{lineno}  
\usepackage{xspace} 
\usepackage[bottom]{footmisc}

\usepackage{graphicx}  
\usepackage{color}
\usepackage{colortbl}
\usepackage{overpic}
\graphicspath{{./figs/}} 

\usepackage{amsmath} 
\usepackage{amssymb}
\usepackage{amsfonts}
\usepackage{upgreek} 

\newcommand*\patchAmsMathEnvironmentForLineno[1]{%
\expandafter\let\csname old#1\expandafter\endcsname\csname #1\endcsname
\expandafter\let\csname oldend#1\expandafter\endcsname\csname
end#1\endcsname
 \renewenvironment{#1}%
   {\linenomath\csname old#1\endcsname}%
   {\csname oldend#1\endcsname\endlinenomath}%
}
\newcommand*\patchBothAmsMathEnvironmentsForLineno[1]{%
  \patchAmsMathEnvironmentForLineno{#1}%
  \patchAmsMathEnvironmentForLineno{#1*}%
}
\AtBeginDocument{%
\patchBothAmsMathEnvironmentsForLineno{equation}%
\patchBothAmsMathEnvironmentsForLineno{align}%
\patchBothAmsMathEnvironmentsForLineno{flalign}%
\patchBothAmsMathEnvironmentsForLineno{alignat}%
\patchBothAmsMathEnvironmentsForLineno{gather}%
\patchBothAmsMathEnvironmentsForLineno{multline}%
}

\usepackage{hyperref}    
\usepackage[all]{hypcap} 

\def\lhcb {\mbox{LHCb}\xspace}
\def\ux85 {\mbox{UX85}\xspace}

\ifthenelse{\boolean{uprightparticles}}%
{

 \def\Ppi         {\ensuremath{\uppi}\xspace}

 \def\PDelta      {\ensuremath{\Delta}\xspace}                 
 \def\PXi      {\ensuremath{\Xi}\xspace}                 
 \def\PLambda      {\ensuremath{\Lambda}\xspace}                 
 \def\PSigma      {\ensuremath{\Sigma}\xspace}                 
 \def\POmega      {\ensuremath{\Omega}\xspace}                 
 \def\PUpsilon      {\ensuremath{\Upsilon}\xspace}                 
 
 \def\PB      {\ensuremath{\mathrm{B}}\xspace}                 
                  
 \def\PD      {\ensuremath{\mathrm{D}}\xspace}

 \def\PK      {\ensuremath{\mathrm{K}}\xspace}

 \def\Pb      {\ensuremath{\mathrm{b}}\xspace}                 
 \def\Pc      {\ensuremath{\mathrm{c}}\xspace}

 \def\Pi      {\ensuremath{\mathrm{i}}\xspace}

 \def\Pp      {\ensuremath{\mathrm{p}}\xspace}

 \def\Ps      {\ensuremath{\mathrm{s}}\xspace}                 
                  
 \def\Pu      {\ensuremath{\mathrm{u}}\xspace}

}
{

 \def\Ppi         {\ensuremath{\pi}\xspace}

 \mathchardef\PDelta="7101
 \mathchardef\PXi="7104
 \mathchardef\PLambda="7103
 \mathchardef\PSigma="7106
 \mathchardef\POmega="710A
 \mathchardef\PUpsilon="7107
                  
 \def\PB      {\ensuremath{B}\xspace}                 
                  
 \def\PD      {\ensuremath{D}\xspace}

 \def\PK      {\ensuremath{K}\xspace}

 \def\Pb      {\ensuremath{b}\xspace}                 
 \def\Pc      {\ensuremath{c}\xspace}

 \def\Pi      {\ensuremath{i}\xspace}

 \def\Pp      {\ensuremath{p}\xspace}

 \def\Ps      {\ensuremath{s}\xspace}                 
                  
 \def\Pu      {\ensuremath{u}\xspace}

}

\def\uquark    {\ensuremath{\Pu}\xspace}

\def\squark    {\ensuremath{\Ps}\xspace}

\def\cquark    {\ensuremath{\Pc}\xspace}

\def\bquark    {\ensuremath{\Pb}\xspace}

\def\pion  {\ensuremath{\Ppi}\xspace}
\def\piz   {\ensuremath{\pion^0}\xspace}

\def\pip   {\ensuremath{\pion^+}\xspace}
\def\pim   {\ensuremath{\pion^-}\xspace}

\def\kaon  {\ensuremath{\PK}\xspace}
  \def\Kbar  {\kern 0.2em\overline{\kern -0.2em \PK}{}\xspace}

\def\Kz    {\ensuremath{\kaon^0}\xspace}
\def\Kzb   {\ensuremath{\Kbar^0}\xspace}
\def\KzKzb {\ensuremath{\Kz \kern -0.16em \Kzb}\xspace}
\def\Kp    {\ensuremath{\kaon^+}\xspace}
\def\Km    {\ensuremath{\kaon^-}\xspace}

\def\KpKm  {\ensuremath{\Kp \kern -0.16em \Km}\xspace}

\def\Kstarz  {\ensuremath{\kaon^{*0}}\xspace}
\def\Kstarzb {\ensuremath{\Kbar^{*0}}\xspace}

  \def\Dbar    {\kern 0.2em\overline{\kern -0.2em \PD}{}\xspace}
\def\D       {\ensuremath{\PD}\xspace}

\def\Dz      {\ensuremath{\D^0}\xspace}
\def\Dzb     {\ensuremath{\Dbar^0}\xspace}
\def\DzDzb   {\ensuremath{\Dz {\kern -0.16em \Dzb}}\xspace}
\def\Dp      {\ensuremath{\D^+}\xspace}
\def\Dm      {\ensuremath{\D^-}\xspace}

\def\DpDm    {\ensuremath{\Dp {\kern -0.16em \Dm}}\xspace}
\def\Dstar   {\ensuremath{\D^*}\xspace}

\def\Dstarz  {\ensuremath{\D^{*0}}\xspace}
\def\Dstarzb {\ensuremath{\Dbar^{*0}}\xspace}
\def\Dstarp  {\ensuremath{\D^{*+}}\xspace}

\def\Ds      {\ensuremath{\D^+_\squark}\xspace}

\def\B       {\ensuremath{\PB}\xspace}
  \def\Bbar    {\kern 0.18em\overline{\kern -0.18em \PB}{}\xspace}

\def\Bz      {\ensuremath{\B^0}\xspace}
\def\Bzb     {\ensuremath{\Bbar^0}\xspace}
\def\Bu      {\ensuremath{\B^+}\xspace}

\def\Bp      {\ensuremath{\Bu}\xspace}

\def\Bs      {\ensuremath{\B^0_\squark}\xspace}
\def\Bsb     {\ensuremath{\Bbar^0_\squark}\xspace}

  \def\Y#1S{\ensuremath{\PUpsilon{(#1S)}}\xspace}

\def\proton      {\ensuremath{\Pp}\xspace}

\def\L {\ensuremath{\PLambda}\xspace}
\def\Lbar {\ensuremath{\kern 0.1em\overline{\kern -0.1em\PLambda}}\xspace}

\def\Lb      {\ensuremath{\L^0_\bquark}\xspace}

\def\to                 {\ensuremath{\rightarrow}\xspace}

\def\CP                {\ensuremath{C\!P}\xspace}

\def\AT#1     {\ensuremath{A_{\mathrm{T}}^{#1}}\xspace}           

\def\C#1      {\ensuremath{\mathcal{C}_{#1}}\xspace}                       
\def\Cp#1     {\ensuremath{\mathcal{C}_{#1}^{'}}\xspace}                    
\def\Ceff#1   {\ensuremath{\mathcal{C}_{#1}^{\mathrm{(eff)}}}\xspace}        
\def\Cpeff#1  {\ensuremath{\mathcal{C}_{#1}^{'\mathrm{(eff)}}}\xspace}       
\def\Ope#1    {\ensuremath{\mathcal{O}_{#1}}\xspace}                       
\def\Opep#1   {\ensuremath{\mathcal{O}_{#1}^{'}}\xspace}                    

\newcommand{\tev}{\ensuremath{\mathrm{\,Te\kern -0.1em V}}\xspace}
\newcommand{\gev}{\ensuremath{\mathrm{\,Ge\kern -0.1em V}}\xspace}
\newcommand{\mev}{\ensuremath{\mathrm{\,Me\kern -0.1em V}}\xspace}
\newcommand{\kev}{\ensuremath{\mathrm{\,ke\kern -0.1em V}}\xspace}
\newcommand{\ev}{\ensuremath{\mathrm{\,e\kern -0.1em V}}\xspace}
\newcommand{\gevc}{\ensuremath{{\mathrm{\,Ge\kern -0.1em V\!/}c}}\xspace}
\newcommand{\mevc}{\ensuremath{{\mathrm{\,Me\kern -0.1em V\!/}c}}\xspace}
\newcommand{\gevcc}{\ensuremath{{\mathrm{\,Ge\kern -0.1em V\!/}c^2}}\xspace}
\newcommand{\gevgevcccc}{\ensuremath{{\mathrm{\,Ge\kern -0.1em V^2\!/}c^4}}\xspace}
\newcommand{\mevcc}{\ensuremath{{\mathrm{\,Me\kern -0.1em V\!/}c^2}}\xspace}

\def\mum  {\ensuremath{\,\upmu\rm m}\xspace}

\def\invfb   {\ensuremath{\mbox{\,fb}^{-1}}\xspace}

\newcommand{\chisq}{\ensuremath{\chi^2}\xspace}

\def\gsim{{~\raise.15em\hbox{$>$}\kern-.85em
          \lower.35em\hbox{$\sim$}~}\xspace}
\def\lsim{{~\raise.15em\hbox{$<$}\kern-.85em
          \lower.35em\hbox{$\sim$}~}\xspace}

\def\PDF {PDF\xspace}

\def\pt         {\mbox{$p_{\rm T}$}\xspace}

\def\dllkpi     {\ensuremath{\mathrm{DLL}_{\kaon\pion}}\xspace}

\def\mrad{\ensuremath{\rm \,mrad}\xspace}

\def\evtgen     {\mbox{\textsc{EvtGen}}\xspace}
\def\pythia     {\mbox{\textsc{Pythia}}\xspace}

\def\geant      {\mbox{\textsc{Geant4}}\xspace}

\def\tell1  {TELL1\xspace}
\def\ukl1   {UKL1\xspace}

\newcommand{\eg}{\mbox{\itshape e.g.}\xspace}
\newcommand{\ie}{\mbox{\itshape i.e.}}

\def\parenbar#1{{\null\!                        
   \mathop#1\limits^{\hbox{\tiny (---)}}       	
   \!\null}}                                    
\def\BzBsbar    {\ensuremath{\parenbar{\B}{^0_{\mskip -3mu (s)}}}\xspace}

\usepackage{cite} 
\usepackage{mciteplus}

\begin{document}

\renewcommand{\thefootnote}{\fnsymbol{footnote}}
\setcounter{footnote}{1}


\begin{titlepage}
\pagenumbering{roman}

\vspace*{-1.5cm}
\centerline{\large EUROPEAN ORGANIZATION FOR NUCLEAR RESEARCH (CERN)}
\vspace*{1.5cm}
\hspace*{-0.5cm}
\begin{tabular*}{\linewidth}{lc@{\extracolsep{\fill}}r}
\ifthenelse{\boolean{pdflatex}}
{\vspace*{-2.7cm}\mbox{\!\!\!\includegraphics[width=.14\textwidth]{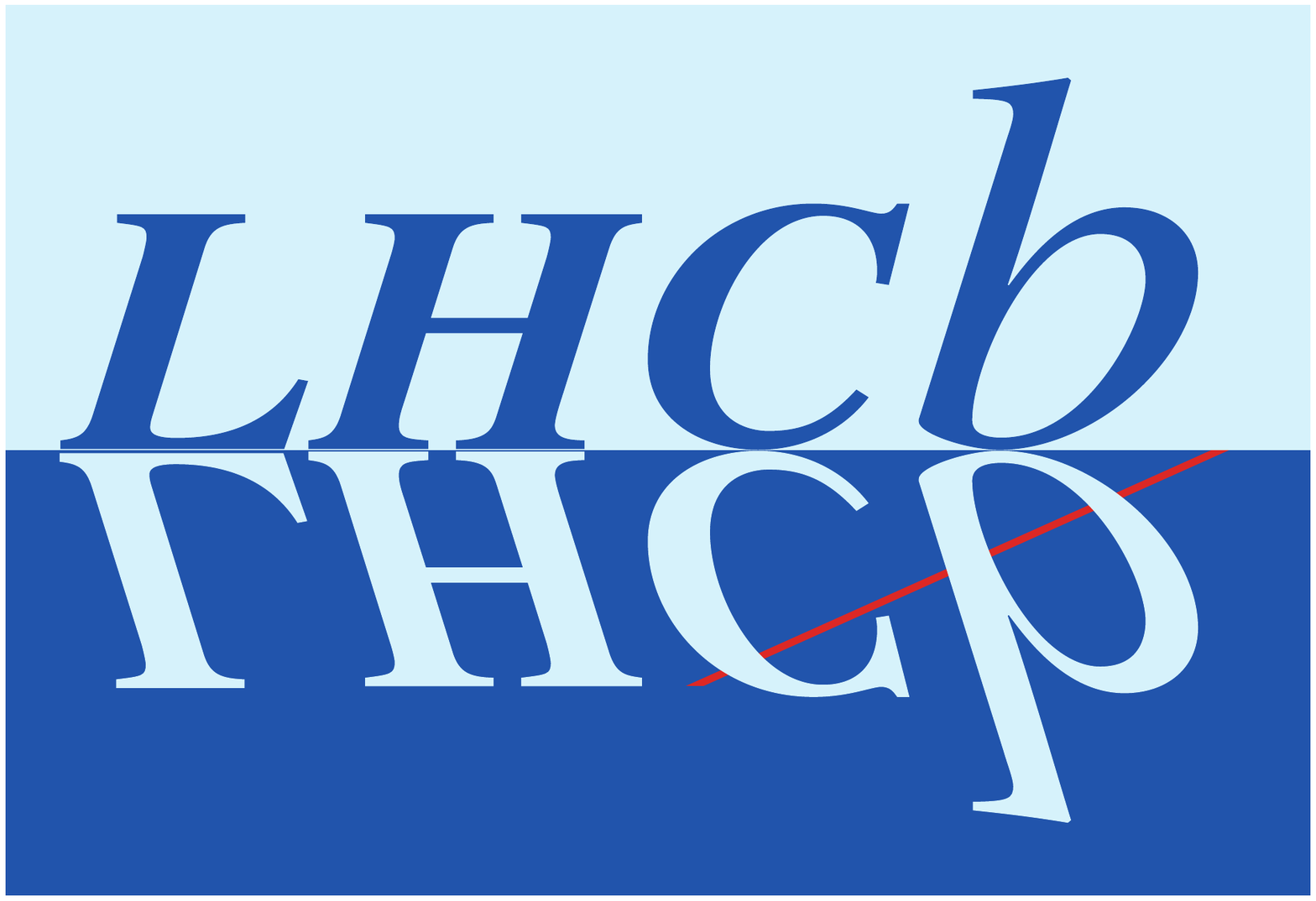}} & &}%
{\vspace*{-1.2cm}\mbox{\!\!\!\includegraphics[width=.12\textwidth]{lhcb-logo.eps}} & &}%
\\
 & & CERN-PH-EP-2012-362 \\  
 & & LHCb-PAPER-2012-042 \\  

& & December 18, 2012 \\
& & \\

\end{tabular*}

\vspace*{2.0cm}

{\bf\boldmath\huge
\begin{center}
  Measurement of \CP observables in $\Bz \to \D \Kstarz$
with $\D\to \Kp\Km$
\end{center}
}

\vspace*{1.0cm}

\begin{center}
The \lhcb collaboration\footnote{Authors are listed on the following pages.}
\end{center}

\vspace{\fill}

\begin{abstract}
  \noindent
The decay  $\Bz \to \D \Kstarz$ and the charge conjugate mode
are studied using 1.0\invfb of $pp$ collision data collected by the \lhcb experiment at $\sqrt{s} = 7\tev$ in 2011.
The \CP asymmetry between the $\Bz \to \D \Kstarz$ and the $\Bzb \to \D \Kstarzb$ decay rates, 
with the neutral \D meson in the \CP-even final state $\Kp\Km$,  is found to be 
\begin{equation*}
{\cal A}_d^{KK}  =  -0.45 \pm 0.23 \pm 0.02,
\end{equation*}
where the first uncertainty is statistical and the
second is systematic.
In addition, favoured $\Bz \to \D \Kstarz$ decays are reconstructed 
with the $D$ meson in the non-$CP$ eigenstate $\Kp\pim$. 
The ratio of the $B$-flavour averaged decay rates in $D$ decays to $CP$ and non-$CP$ eigenstates
is measured to be 
\begin{equation*}
{\cal R}_d^{KK}  =  1.36\,^{+\,0.37}_{ -\,0.32} \pm 0.07,
\end{equation*}
where the ratio of the branching fractions of $\Dz \to \Km\pip$ to $\Dz\to \Kp\Km$ decays is included as multiplicative factor.
The \CP asymmetries measured with two control channels, 
the favoured $\Bz\to\D\Kstarz$ decay with $D\to \Kp\pim$ and the $\Bsb\to \D \Kstarz$ decay with $\D\to \Kp\Km$, are also reported.
\end{abstract}

\vspace*{1.0cm}

\begin{center}
  Submitted to JHEP
\end{center}

\vspace{\fill}

{\footnotesize 
\centerline{\copyright~CERN on behalf of the \lhcb collaboration, license \href{http://creativecommons.org/licenses/by/3.0/}{CC-BY-3.0}.}}
\vspace*{2mm}

\end{titlepage}


\newpage
\setcounter{page}{2}
\mbox{~}
\newpage

\centerline{\large\bf LHCb collaboration}
\begin{flushleft}
\small
R.~Aaij$^{38}$, 
C.~Abellan~Beteta$^{33,n}$, 
A.~Adametz$^{11}$, 
B.~Adeva$^{34}$, 
M.~Adinolfi$^{43}$, 
C.~Adrover$^{6}$, 
A.~Affolder$^{49}$, 
Z.~Ajaltouni$^{5}$, 
J.~Albrecht$^{35}$, 
F.~Alessio$^{35}$
% ...

{\footnotesize \it
% ...
$ ^{5}$Clermont Universit\'{e}, Universit\'{e} Blaise Pascal, CNRS/IN2P3, LPC, Clermont-Ferrand, France\\
$ ^{6}$CPPM, Aix-Marseille Universit\'{e}, CNRS/IN2P3, Marseille, France\\
% ...
$ ^{11}$Physikalisches Institut, Ruprecht-Karls-Universit\"{a}t Heidelberg, Heidelberg, Germany\\
% ...
$ ^{33}$Universitat de Barcelona, Barcelona, Spain\\
$ ^{34}$Universidad de Santiago de Compostela, Santiago de Compostela, Spain\\
$ ^{35}$European Organization for Nuclear Research (CERN), Geneva, Switzerland\\
% ...
$ ^{38}$Nikhef National Institute for Subatomic Physics, Amsterdam, The Netherlands\\
% ...
$ ^{43}$H.H. Wills Physics Laboratory, University of Bristol, Bristol, United Kingdom\\
% ...
$ ^{49}$Oliver Lodge Laboratory, University of Liverpool, Liverpool, United Kingdom\\
% ...
$ ^{n}$LIFAELS, La Salle, Universitat Ramon Llull, Barcelona, Spain\\
% ...
}
\end{flushleft}

\cleardoublepage


\renewcommand{\thefootnote}{\arabic{footnote}}
\setcounter{footnote}{0}



\pagestyle{plain} 
\setcounter{page}{1}
\pagenumbering{arabic}


%


\section{Introduction}\label{Introduction}

\begin{figure}[!b]
\centering
\begin{overpic}[width=0.49\textwidth]{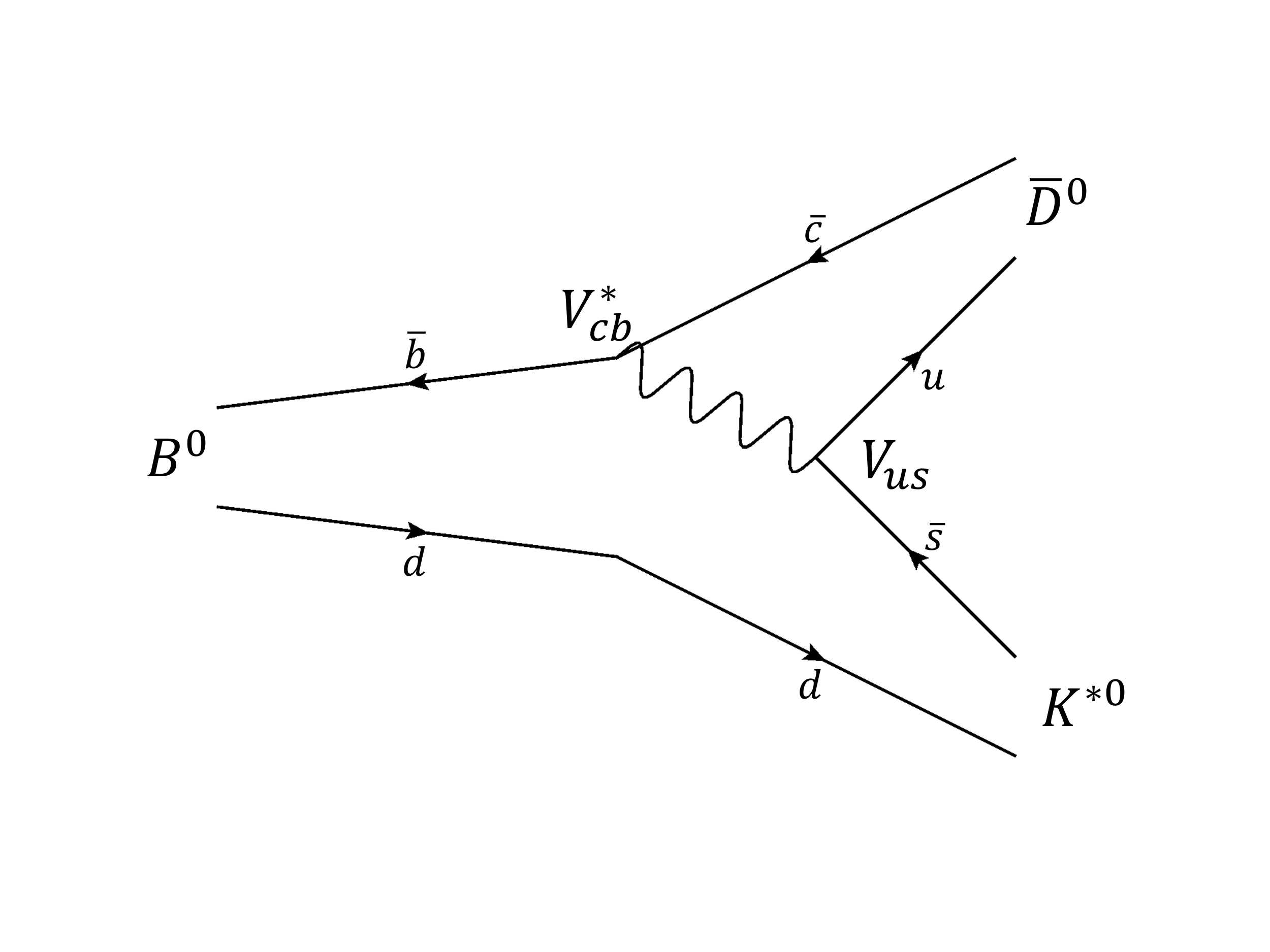}
\put(10,60){(a)} 
\end{overpic}
\begin{overpic}[width=0.49\textwidth]{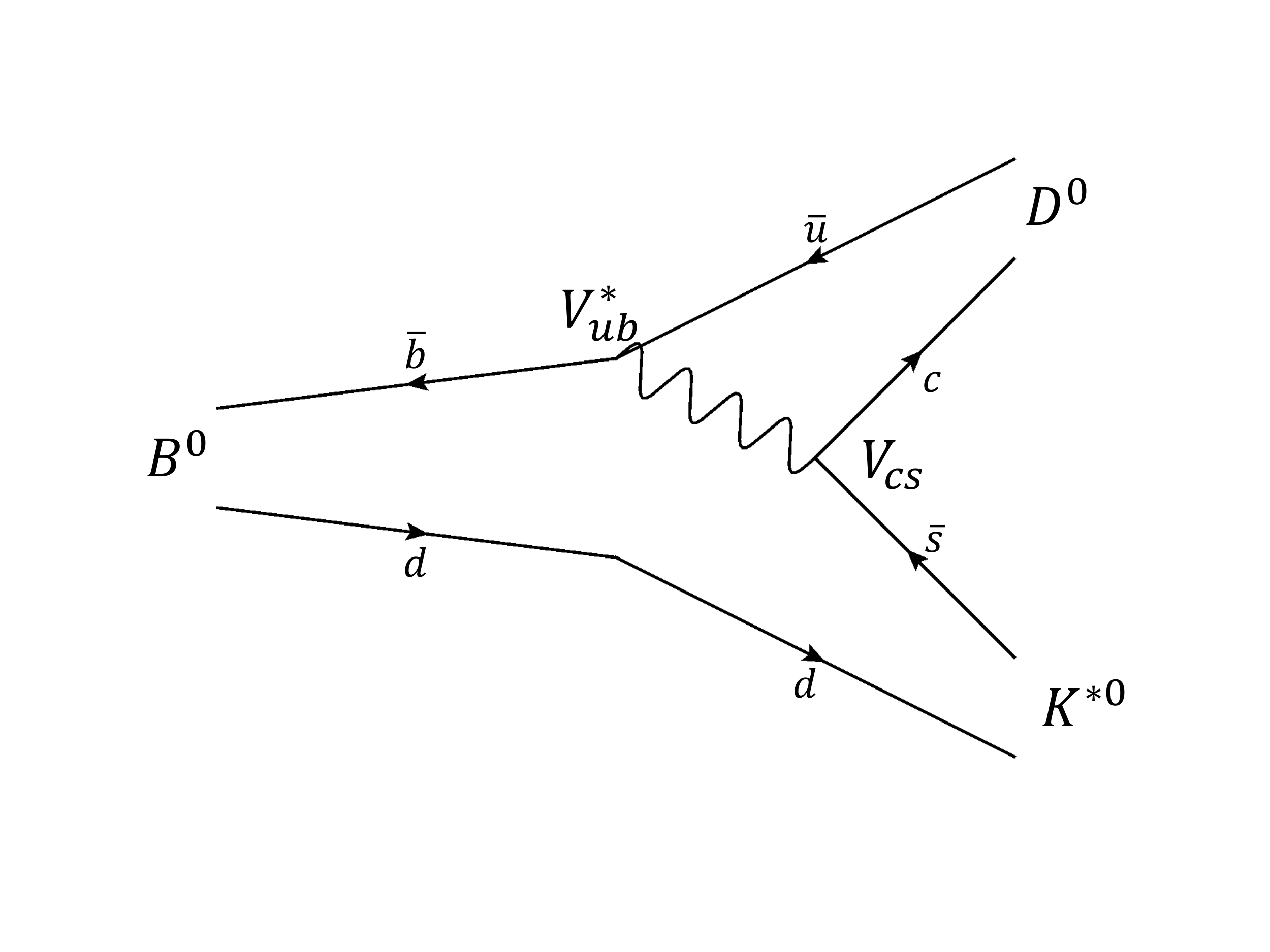}
\put(10,60){(b)} 
\end{overpic}
\caption{\small Feynman diagrams for (a) $\Bz\to \Dzb \Kstarz$  and  (b) $\Bz\to \Dz \Kstarz$.}
\label{fig:Feynman}
\end{figure}

Direct \CP violation can arise in $\Bz\to \D \Kstarz$ 
decays\footnote{Here and in the following, \D represents a neutral meson that is an admixture of \Dz and \Dzb. Inclusion of 
charge conjugate modes is implied unless specified otherwise.} from the interference between two colour-suppressed
transitions: $\bar\bquark\to \bar\cquark$ (Cabibbo favoured) and $\bar\bquark\to \bar\uquark$ (Cabibbo suppressed). The corresponding  
Feynman diagrams are shown in Fig.~\ref{fig:Feynman}; interference occurs if the \Dz and \Dzb mesons decay to a common final state.
The magnitude of the \CP-violating asymmetry that arises from this interference is related to the value of the phase $\gamma = \arg\left[- (V_{ud}V_{ub}^*)/(V_{cd}V_{cb}^*)\right]$, 
the least-well determined angle of the Unitarity Triangle. 
A method to determine $\gamma$ from hadronic \B-decay rates was originally proposed by Gronau, London and Wyler (GLW)
in Ref.~\cite{bib:GL,*bib:GW} for various charged and neutral $\B\to \D\kaon$ decay modes 
and can be applied to the decay mode $\Bz\to \D\Kstarz$. 
In this mode, the charge of the kaon from the $\Kstarz\to\Kp\pim$ decay 
unambiguously tags the flavour of the decaying \B meson~\cite{Dunietz:1991yd}, hence no time-dependent tagged
analysis is required. 

The use of these neutral $B$ decays is particularly interesting because
the magnitude of the ratio of the suppressed over the favoured amplitude, which controls the size 
of the interference, is expected to be relatively large 
(naively a factor three larger than the analogous ratio for $\Bp\to D\Kp$ decays),
hence the system can exhibit large \CP-violating effects, depending on the $D$ decay. 
Among the modes used in the GLW method, which are studied in this paper, large \CP asymmetries can be 
expected when the $D$ meson is reconstructed in a \CP eigenstate.
Contributions from \Bz decays to the non-resonant $\D\Kp\pim$ final state, which can pollute the 
$\D\Kstarz$ reconstructed signal candidates due to the large
natural width of the \Kstarz, can be treated in a model-independent way, as shown in Ref.~\cite{Gronau:2002mu}.
Studies with simulated events have shown that the $\Bz\to \D\Kstarz$ mode is one of the most promising channels to provide
a precise measurement of $\gamma$ at \lhcb~\cite{Adeva:2009ny}. Results with this channel will therefore complement those 
from $\Bp\to D\Kp$, which have recently been used by LHCb to constrain $\gamma$~\cite{Aaij:2012kz, *Aaij:1483187}.

This paper presents the measurement of the $\Bz-\Bzb$ partial width asymmetry using 
\D decays into the \CP eigenstate $\Kp\Km$,
\begin{equation}\label{eq:akk}
{\cal A}_d^{KK} = \frac{\Gamma(\Bzb \to \D_{[\Kp\Km]} \Kstarzb) - \Gamma(\Bz \to \D_{[\Kp\Km]} \Kstarz)}{\Gamma(\Bzb \to D_{[\Kp\Km]} \Kstarzb) + \Gamma(\Bz \to \D_{[\Kp\Km]} \Kstarz)},
\end{equation}
together with the measurement of the ratio of the average of the \Bz and \Bzb partial widths 
with $\D\to \Kp\Km$, to the average partial width with $\D \to \Kp \pim$ 
(where the sign of the kaon charge from the \D decay is the same as that 
of the kaon from the \Kstarz decay),
\begin{equation}\label{eq:rpipi}
{\cal R}_d^{KK}  =  \frac{\Gamma(\Bzb \to \D_{[\Kp\Km]} \Kstarzb) + \Gamma(\Bz \to \D_{[\Kp\Km]} \Kstarz)}{\Gamma(\Bzb \to \D_{[\Km\pip]} \Kstarzb) + \Gamma(\Bz \to \D_{[\Kp\pim]} \Kstarz)}.
\end{equation}
These quantities can be used together with other inputs to determine the value of $\gamma$. Note that the suppressed decay mode
$\Bz \to \D_{[\Km\pip]} \Kstarz$, where the sign of the kaon charge from the \D decay is opposite to that 
of the kaon from the \Kstarz decay, is not included in this analysis. This decay mode can exhibit large \CP-violating effects and can be studied
with a larger dataset.
The measured asymmetry in the favoured decay $\Bz\to\D_{[\Kp\pim]}\Kstarz$,
\begin{equation}\label{eq:afav}
{\cal A}_d^{\rm fav} = \frac{\Gamma(\Bzb \to \D_{[\Km\pip]} \Kstarzb) - \Gamma(\Bz \to \D_{[\Kp\pim]} \Kstarz)}{\Gamma(\Bzb \to \D_{[\Km\pip]} \Kstarzb) + \Gamma(\Bz \to \D_{[\Kp\pim]} \Kstarz)}
\end{equation}
is a useful cross-check since it is expected to be compatible with zero given the size of the current dataset.

In $pp$ collisions, \Bs mesons are produced and can decay to the same final state, 
$\Bs\to \D \Kstarzb$~\cite{LHCb-PAPER-2011-008}. In these \Bs decay modes, the interference between the two contributing amplitudes is 
expected to be small, since the relative magnitude of the suppressed to the favoured amplitude is 
small compared to the \Bz case.
Therefore, these modes are valuable control channels, and the asymmetry 
\begin{equation}\label{eq:as}
{\cal A}^{KK}_s = \frac{\Gamma(\Bsb \to \D_{[\Kp\Km]} \Kstarz) - \Gamma(\Bs \to \D_{[\Kp\Km]} \Kstarzb)}{\Gamma(\Bsb \to \D_{[\Kp\Km]} \Kstarz) + \Gamma(\Bs \to \D_{[\Kp\Km]} \Kstarzb)},
\end{equation}
similar to that defined in 
Eq.~\eqref{eq:akk}, is also obtained in this analysis. Since the favoured (suppressed) \Bsb (\Bz) decay gives kaons with opposite charges from \D and \Kstarz decays, ${\cal A}_s^{\rm fav}$ is not used as a control measurement in the analysis, to avoid biasing a potential future measurement of ${\cal A}_d^{\rm sup}$.

\section{The \lhcb detector, dataset and event selection}\label{detector}

The study reported here is based on a data sample collected at the Large Hadron Collider (LHC) with the LHCb 
detector at a centre-of-mass energy of 7\tev during the year 2011,
corresponding to an integrated luminosity of 1.0\invfb.
The \lhcb detector~\cite{Alves:2008zz} is a single-arm forward
spectrometer covering the \mbox{pseudorapidity} range $2<\eta <5$,
designed for the study of particles containing \bquark or \cquark
quarks. The detector includes a high precision tracking system
consisting of a silicon-strip vertex detector surrounding the $pp$
interaction region, a large-area silicon-strip detector located
upstream of a dipole magnet with a bending power of about
$4{\rm\,Tm}$, and three stations of silicon-strip detectors and straw
drift tubes placed downstream. The combined tracking system has a
momentum resolution $\Delta p/p$ that varies from 0.4\% at 5\gevc to
0.6\% at 100\gevc, and an impact parameter resolution of 20\mum for
tracks with high transverse momentum (\pt). Charged hadrons are identified
using two ring-imaging Cherenkov detectors. Photon, electron and
hadron candidates are identified by a calorimeter system consisting of
scintillating-pad and preshower detectors, an electromagnetic
calorimeter and a hadronic calorimeter. Muons are identified by a
system composed of alternating layers of iron and multiwire
proportional chambers. The trigger~\cite{Aaij:2012me} consists of a hardware stage, based
on information from the calorimeter and muon systems, followed by a
software stage which applies a full event reconstruction.

This analysis uses events selected by the hardware level trigger either when one of the charged particles of the signal
decay gives a large enough energy deposit in the calorimeter system (hadron trigger), or when one of the particles
in the event, not coming from the signal decay, fulfills the trigger requirements (\ie\ mainly events triggered
by one particle coming from the decay of the other \B in the event). The software trigger requires a two-, three- or four-track
secondary vertex with a high scalar sum of the \pt of
the tracks and a significant displacement from the primary $pp$ interaction vertices~(PVs).
At least one track should have $\pt > 1.7\gevc$ and an impact parameter~(IP)
\chisq with respect to the PV greater than 16. The IP \chisq
is defined as the difference between the \chisq of the PV
reconstructed with and without the considered track. A multivariate algorithm is used
for the identification of secondary vertices consistent with the decay of a \bquark hadron.

Candidates are selected from combinations of charged particles. 
$D$ mesons are reconstructed in the decay modes $\D\to \Kp \pim$ and $\Kp\Km$. 
The \pt of the daughters is required to be larger than 400\mevc. 
Particle identification (PID) is used to distinguish between charged pions and kaons. The difference between
the log-likelihoods of the kaon and pion hypotheses (\dllkpi) is required to be larger
than 0 for kaons and smaller than 4 for pions. This aids the reduction of cross-feed between the signal $\D$ decay modes to a negligible level.
A fit is applied to the two-track vertex, requiring 
that the corresponding 
\chisq per degree of freedom is less than 5. In order to separate \D mesons coming from a \B decay from 
those produced at the PV, the \D candidates are required to have an IP \chisq 
greater than 4 with respect to any PV. 
To suppress background from \B decays without an intermediate \D meson
($\Bz\to \Kstarz \Kp\Km$ for example), for which all four charged hadrons are produced at
the \B-decay vertex, a condition on the \D flight distance with respect to the \B vertex is 
applied, requiring
that it is larger than 0 by at least 2.5 standard deviations. Finally, \D candidates with an invariant mass within $\pm 20\mevcc$ 
of the nominal \Dz mass are retained. 

\Kstarz mesons are reconstructed in the mode $\Kstarz \to \Kp \pim$. The \pt of the
\Kp and \pim mesons must be larger than 300\mevc. PID is
also used, requiring that \dllkpi is larger than 3 for the kaon and lower than 3 for the
pion, reducing the cross-feed from $\Bz \to \D\rho^0$ to a manageable level and rejecting non-resonant $\Bz \to \D \Kp\Km$~\cite{
Aaij:1464076}.
Possible contamination from protons in the kaon sample, \eg from $\Lb \to \D \proton \pim$ decays, is reduced by removing kaon candidates 
with a difference between the log-likelihoods of the proton and kaon hypotheses (${\rm DLL}_{pK}$) of 
less than 10.
The IP \chisq of the \Kstarz mesons must be larger than 25, to select those coming from a \B decay, 
and their invariant mass within $\pm 50\mevcc$ of the nominal mass. 

\BzBsbar meson candidates are formed by combining \D and \Kstarz candidates selected with the above
 requirements. A fit to a common vertex is performed, keeping only combinations with \chisq per degree of freedom
 lower than 4, and a kinematic fit is performed to constrain the
invariant mass of the reconstructed \D to the nominal \Dz mass~\cite{Nakamura:2010zzi}. Since \B mesons are produced at the PV, only candidates with IP \chisq
 lower than 9 are retained. In case several PVs are reconstructed, the one for which the \B-candidate IP \chisq 
is the smallest is taken as reference. 
 Additionally, the momentum of the reconstructed
 \B candidate is required to point back to the PV, by requiring that the 
 angle between the \B momentum direction and its direction of flight
 from the PV is smaller than 10\mrad. Furthermore, the sum of the square roots of the IP \chisq
 of the four charged particles must be larger than 32. 
 The absolute value of the cosine of the \Kstarz helicity angle is required to be larger than 0.4. 
 This angle is defined as the angle between the kaon-daughter momentum direction in the \Kstarz rest frame, and the
 \Kstarz direction in the \B rest frame. 
 
 Specific peaking backgrounds from $B^0_{\mskip -3mu (s)} \to D_{\mskip -3mu (s)}^\mp h^\pm$ decays, where $h$ is a \pion or a \kaon meson, are eliminated by vetoing candidates for which the invariant mass of $\Kp\Km\pip$($\Km\pip\pip$ and $\Kp\Km\pip$) is within
 $\pm 15\mevcc$ of the nominal mass of a \Ds(\Dp) meson.
 
Where possible, data-driven methods are used to determine selection efficiencies and invariant mass distribution shapes. Otherwise, they are determined from fully simulated events. 
The $pp$ collisions are generated using
\pythia~6.4~\cite{Sjostrand:2006za} with a specific \lhcb
configuration~\cite{LHCb-PROC-2010-056} where, in particular, decays of hadronic particles
are described by \evtgen~\cite{Lange:2001uf}. The
interaction of the generated particles with the detector and its
response are implemented using the \geant
toolkit~\cite{Allison:2006ve, *Agostinelli:2002hh} as described in
Ref.~\cite{LHCb-PROC-2011-006}.

\section{Determination of signal yields}

The numbers of reconstructed signal \Bz and \Bs candidates are determined from an unbinned maximum 
likelihood fit to their mass distributions.  
Candidates are split into four categories, which are fitted simultaneously:
$\D(\Kp\Km) \Kstarz$, 
$\D(\Kp\Km) \Kstarzb$,
$\D(\Kp\pim) \Kstarz$, and
$\D(\Km\pip) \Kstarzb$.
The mass distribution of each category is fitted with a sum of probability density functions 
(\PDF) modelling the different contributing components:

\begin{enumerate}

\item the \Bz and \Bs signals are described by double Gaussian functions;

\item the combinatorial background is described by an exponential function;

\item the cross-feed from $\Bz\to \D \rho^0$ decays, where one pion from the $\rho^0\to\pip\pim$ decay
is misidentified as a kaon, is described by a non-parametric \PDF~\cite{Cranmer:2000du} determined from fully simulated 
and selected events;

\item the partially reconstructed $\Bz\to \Dstar\Kstarz$ and $\Bsb \to \Dstar \Kstarz$ 
decays, where the \Dstar is a \Dstarz or a \Dstarzb and the \piz or photon from the \Dstar decay is not reconstructed, 
are modelled by a non-parametric \PDF determined from fully simulated and selected events.

\end{enumerate}

There are 23 free parameters in the fit. These include the \Bz PDF peak position, the core Gaussian resolution for 
the \Bz and the \Bs and the slope of the combinatorial background, all of which are common to the four fit categories. The remaining
free parameters are yields for each fit component within each category.
Yields for $\Bz_{\mskip -3mu (s)}$ and  $\Bzb_{\mskip -3mu (s)}$
are constrained to be identical for the background components where \CP violation effects can be excluded or 
are expected to be compatible with zero with the current data sample size.

A separate fit to $\Bz\to \D(\Kp\pim)\rho^0$ candidates in the same data sample is performed. The yield of such candidates 
and the probability to reconstruct them as $\Bz\to \D(\Kp\pim)\Kstarz$ is used
to constrain the number of cross-feed events in the $\D(\Kp\pim) \Kstarz$ category. 
The number of cross-feed candidates from $\Bz\to \D(\Kp\Km)\rho^0$ in the $\D(\Kp \Km) \Kstarz$ category is derived from the $\D(\Kp\pim) \Kstarz$ category 
using the relative \D branching fractions 
and \B selection efficiencies. 
As no flavour asymmetry is expected for this background, 
the numbers of cross-feed events in the $\D\Kstarzb$ categories are constrained to be identical 
to those of the corresponding $\D\Kstarz$ categories.

The partially reconstructed background component accumulates at masses lower than the nominal \Bz mass. 
Its shape depends on the unknown fraction of transverse polarisation in the $\BzBsbar\to \Dstar \Kstarz$ decays. 
In order to model the $\BzBsbar \to \Dstar\Kstarz$ contribution, 
a \PDF is built from a linear combination of three non-parametric functions corresponding to the three orthogonal 
helicity eigenstates. The functions are derived from simulated $\BzBsbar \to \Dstar\Kstarz$ events reconstructed as $\Bz \to \D \Kstarz$. 
Each function corresponds to the weighted sum of the $\Dstar\to \D\gamma$ and $\Dstar\to \D\piz$ contributions 
for a defined helicity eigenstate, where the weights take into account the relative \Dstar decay branching fractions and 
the corresponding reconstruction efficiencies.

\begin{figure}[!b]
\centering
\includegraphics [width=0.495\textwidth]{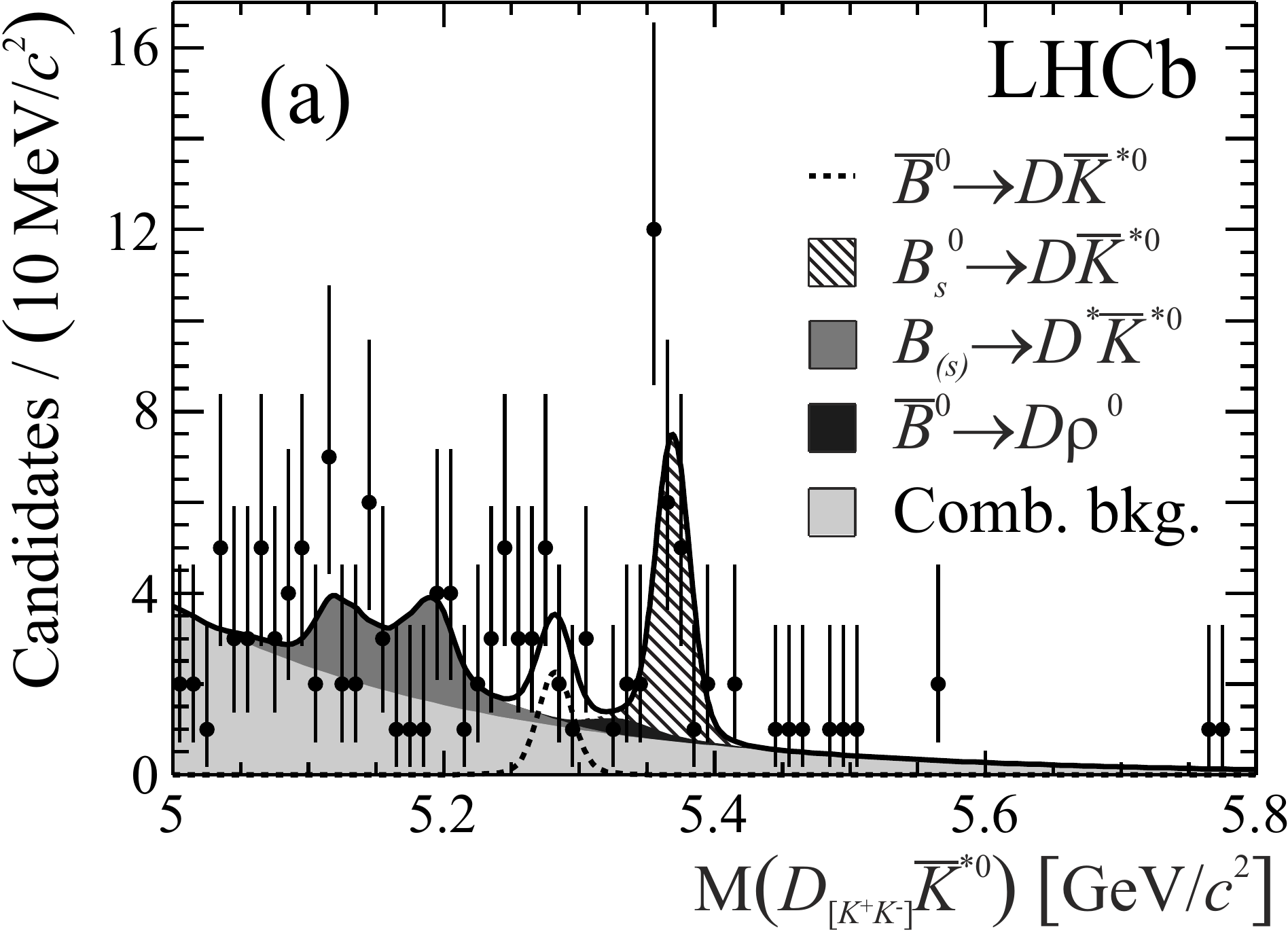}
\includegraphics [width=0.495\textwidth]{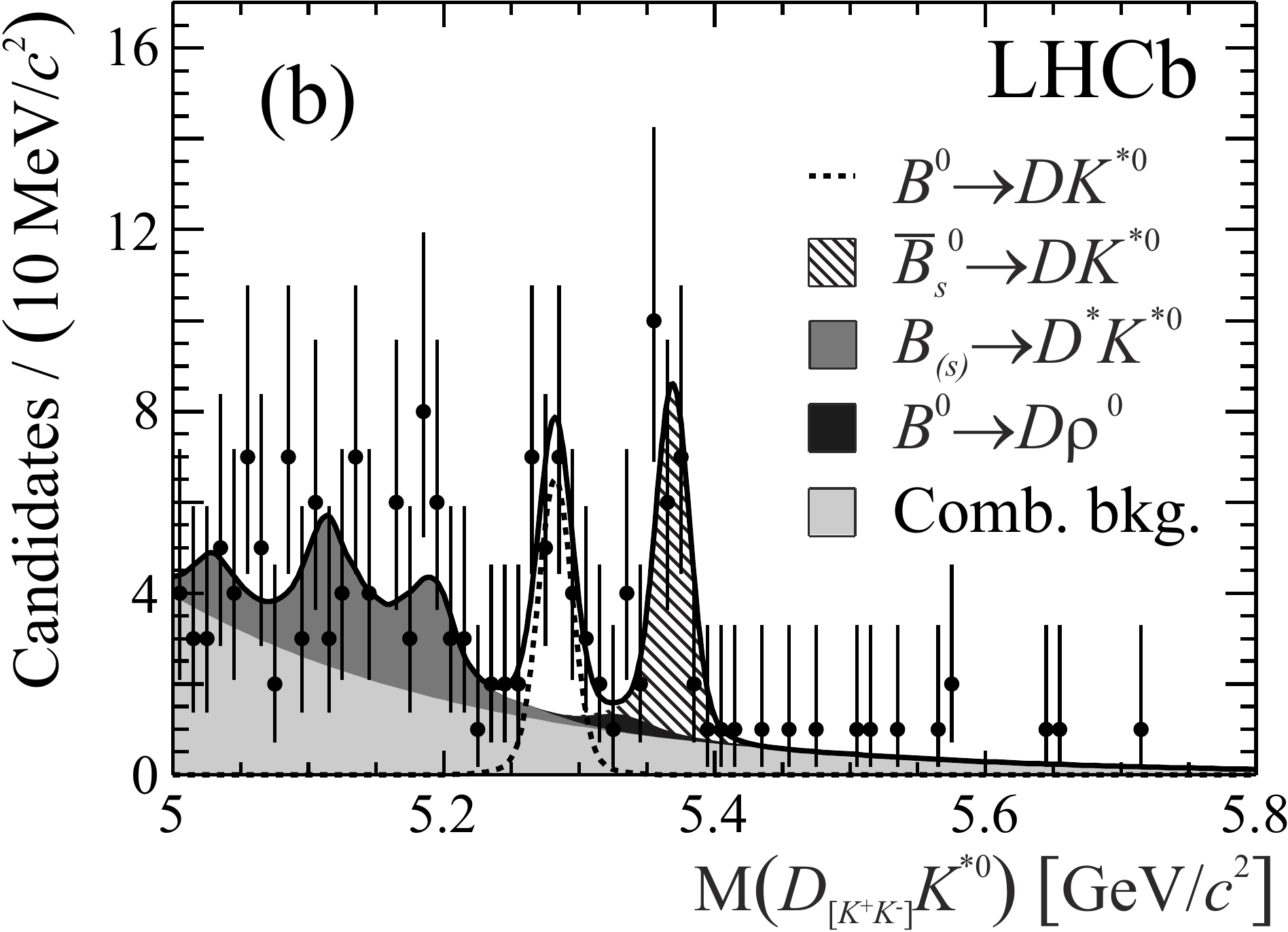}
\includegraphics [width=0.495\textwidth]{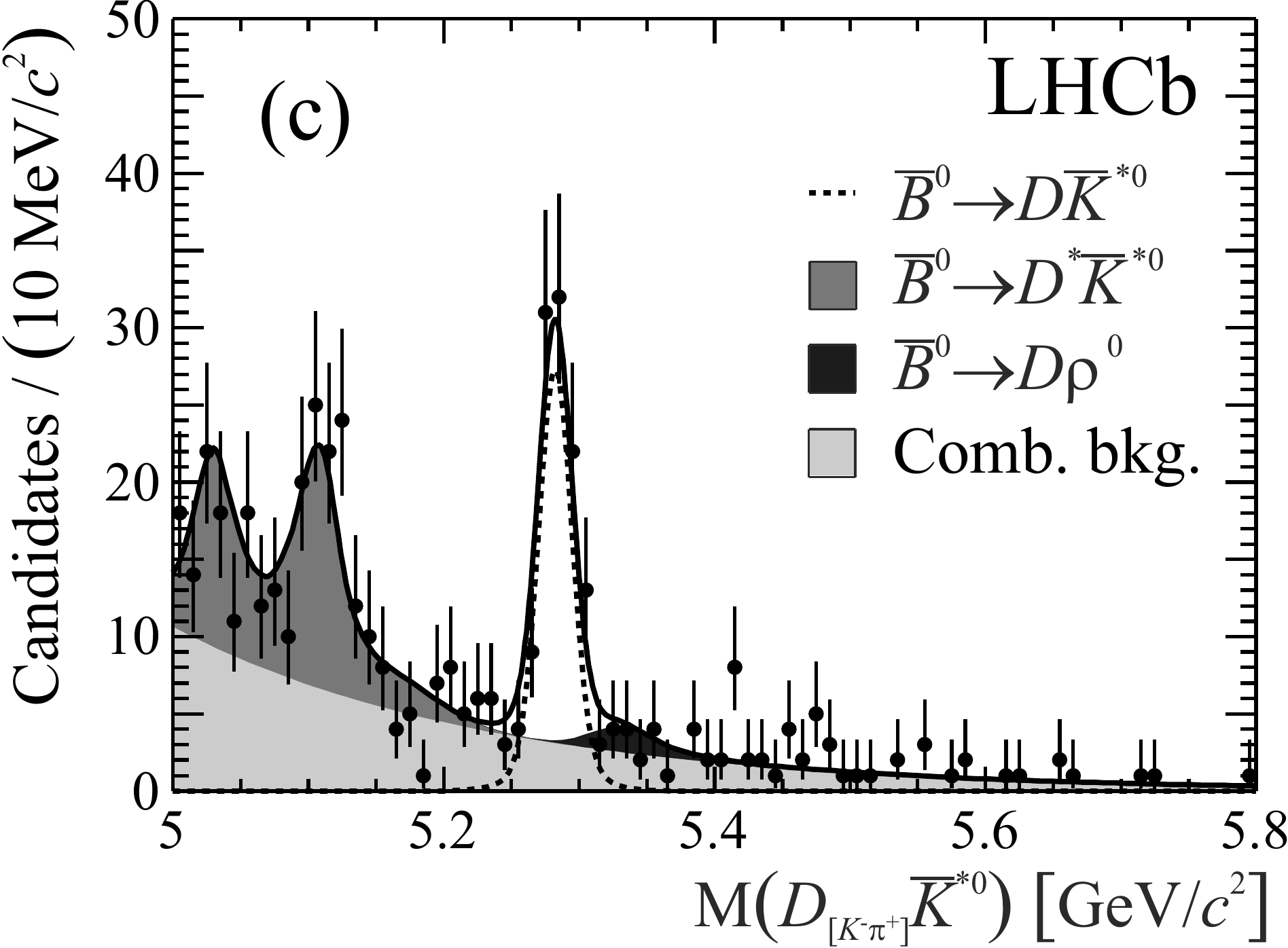}
\includegraphics [width=0.495\textwidth]{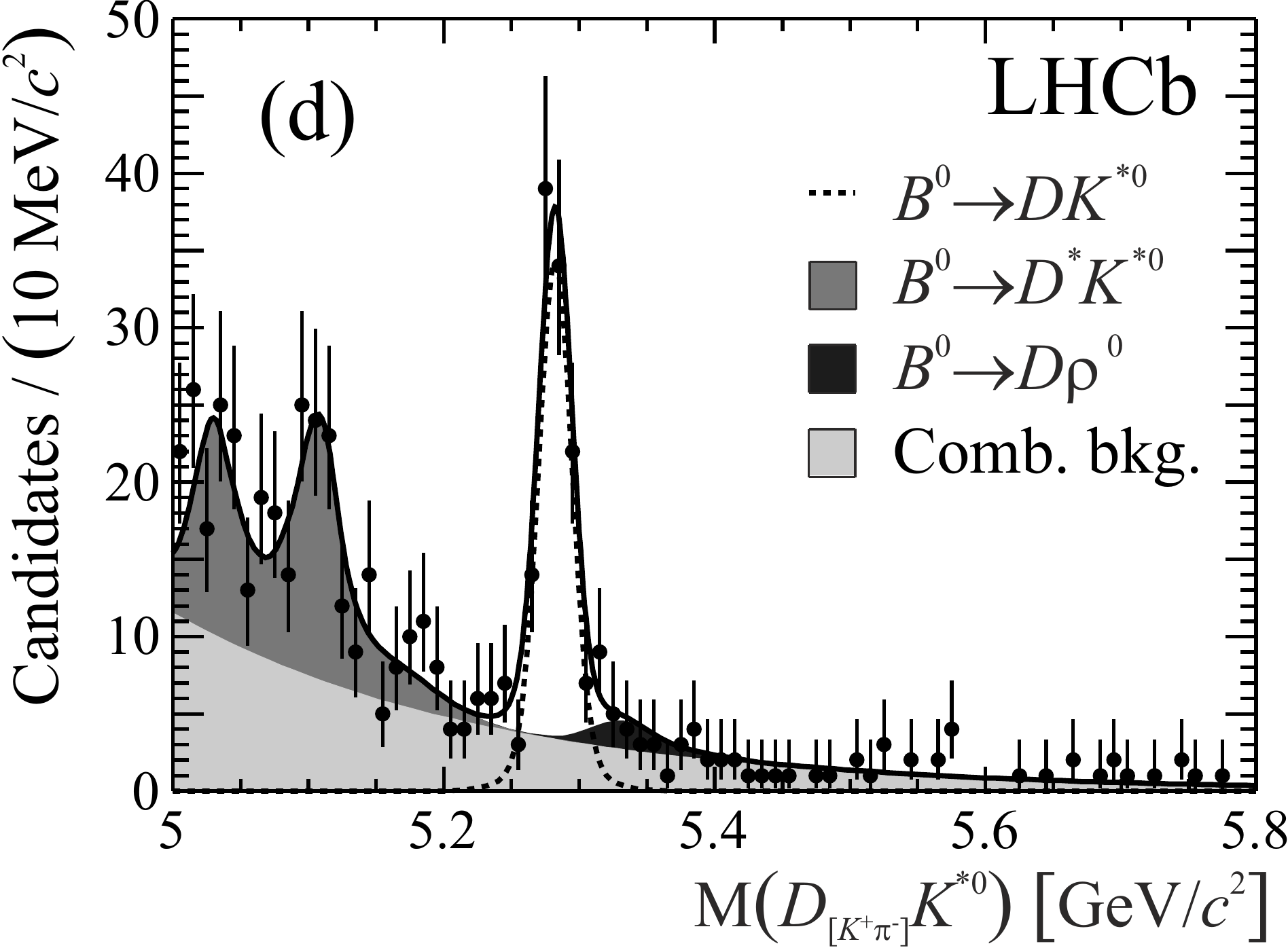}
\caption{\small Invariant mass distributions of ({a}) $\D_{[\Kp\Km]} \Kstarzb$,
({b}) $\D_{[\Kp\Km]} \Kstarz$,
({c}) $\D_{[\Km\pip]} \Kstarzb$ and 
({d}) $\D_{[\Kp\pim]} \Kstarz$ candidates.  
The $\D\Kstarzb$ distributions correspond
to \Bzb and \Bs decays whereas the $\D\Kstarz$ distributions correspond to 
\Bz and \Bsb decays. 
The fit functions are superimposed;
the different \B decays and combinatorial background components are detailed in the legends.}
\label{fig:KK}
\end{figure}

The invariant mass distributions together with the function resulting from the fit are shown in 
Fig.~\ref{fig:KK}. 
Note that the decay $\Bsb\to D(\Kp\pim)\Kstarz$ is not observed since the charge combination of the kaons in the final state
corresponds to the suppressed decay.
The signal yield in each category is
summarized in Table~\ref{tab:fitresult}. The significance of the $\Bz\to \D \Kstarz$ signal
for $\D\to \Kp\Km$ decays, summing \Bz and \Bzb and including both statistical and 
systematic uncertainties, is found to be equal to $5.1\,\sigma$, by comparing the maximum
of the likelihood of the nominal fit and the maximum with the yield of the $\Bz\to \D(\Kp\Km) \Kstarz$ 
category set to zero.

The yields determined from the simultaneous mass fit are corrected for selection efficiencies in order to 
evaluate the asymmetries and ratios described in the introduction. The selection efficiencies account for
the geometrical acceptance of the detector, the reconstruction, the PID, and the trigger efficiencies. All efficiencies
are computed from fully simulated events, except for the PID and trigger efficiencies, which  
are obtained directly from data using clean calibration samples of $\Dz\to \Km\pip$ from \Dstarp 
decays. 

\renewcommand{\arraystretch}{1.5}
\begin{table}[!b]
\tabcolsep 4mm
\begin{center}
\caption{\small \label{tab:fitresult}Signal yields with their statistical uncertainties.}
\begin{tabular}{@{}lr@{}llr@{}l@{}}

Category & \multicolumn{2}{c}{Signal yield} & Category & \multicolumn{2}{c}{Signal yield} \\\midrule
$\Bz\to \D_{[\Kp\Km]} \Kstarz$   & $21\,$ & $_{-\,5}^{+\,6}$ & 
$\Bzb\to \D_{[\Kp\Km]} \Kstarzb$ & $8\,$ & $\pm \, 4$ \\

$\Bs\to \D_{[\Kp\Km]} \Kstarzb$ &  $23\,$ & $_{-\,5}^{+\,6}$ & 
$\Bsb\to \D_{[\Kp\Km]} \Kstarz$ & $24\,$ & $_{\,-\,\,\,5}^{\,+\,\,\,6}$ \\

$\Bz\to \D_{[\Kp\pim]} \Kstarz$ & $\phantom{11}108\,$ & $_{-\,11}^{+\,12}$  &  
$\Bzb\to \D_{[\Km\pip]} \Kstarzb$ & $\phantom{111}94\,$ & $\pm\,11$ \\ 

\end{tabular}
\end{center}
\end{table}
\renewcommand{\arraystretch}{1.} 

\section{Systematic uncertainties}

Several sources of systematic uncertainty are considered, affecting either the determination of the signal yields
or the computation of the efficiencies. They are summarized in Table~\ref{tab:systematics}.
In order to take into account the measured difference in the production rate between \Bzb and \Bz, 
the \Bzb yields are multiplied by a correction factor,
\begin{equation}
a_{\rm prod}^d = \frac{1 - \kappa A_{\rm prod}}{1 + \kappa A_{\rm prod}}, 
\end{equation}
where
$A_{\rm prod}= 0.010 \pm 0.013$~\cite{LHCb-PAPER-2011-029} is the asymmetry between \Bzb and \Bz at production in $pp$ collisions,   
and  $\kappa$ is a decay-dependent 
factor, 
$\kappa=\frac{\int_0^{+\infty}e^{-\Gamma t}\cos(\Delta m t)~\epsilon(\Bz\to\D \Kstarz,t)~{\rm d}t}{\int_0^{+\infty}e^{-\Gamma t}~\epsilon(\Bz\to\D \Kstarz,t)~{\rm d}t}$,
which takes into account dilution effects due to the $\Bz-\Bzb$ oscillation frequency, $\Delta m$, and
includes the acceptance as a function of the decay time 
for the reconstructed decay, $\epsilon(\Bz\to\D \Kstarz,t)$.
The value of $\kappa$ is found to be 
$0.46 \pm 0.01$ using fully simulated events and PID efficiencies from calibration samples.
The uncertainty on $a_{\rm prod}^d$ is propagated to the measured observables 
to estimate the systematic uncertainty from the production asymmetry.
Owing to the large \Bs oscillation frequency, the potential production asymmetry of \Bs mesons does 
not significantly affect the measurement presented here and is neglected.

The PID calibration introduces a systematic uncertainty on the calculated PID efficiencies, which propagates to the final results. All PID correction factors are compatible with unity within their uncertainties which are of the order of 1\%.

The systematic uncertainty associated to the trigger is estimated by varying in the simulation the fraction of events triggered by the hadron trigger with respect to the fraction of events triggered by the other \bquark-hadron 
in the event.
Other selection efficiencies cancel in the ratio of yields, except for the efficiencies of the \pt cuts 
on the \D daughters, which are different  between different \D decay modes.
${\cal{R}}_{d}^{KK}$ has to be corrected by a multiplicative factor $0.94 \pm 0.04$, where the statistical uncertainty 
on the correction, which arises from finite simulated sample size, is assigned as systematic uncertainty due to the relative selection efficiencies.

The fit procedure is validated with simulated experiments. A bias of statistical nature, 
owing to the small number of events in the $\Bz\to \D(\Kp\Km)\Kstarz$ channel, is found to be 5\% for \Bz 
and 8\% for \Bzb. The signal yields are corrected for this bias before computing the 
asymmetries and ratios. A systematic uncertainty equal to half the size of the correction has been assigned. 

Simulated experiments are also used to determine the systematic uncertainties due to the low-mass background, 
the $\Bz\to \D \rho^0$ cross-feed, and the signal shape. 
Samples are generated with different values of the polarisation parameters, the cross-feed fraction and
the fixed signal parameters. The corresponding systematic uncertainty
is estimated from the bias in the results obtained by performing the fit described in the previous section to these samples.

\renewcommand{\arraystretch}{1.4}
\begin{table}[!t]
\tabcolsep 5mm
\begin{center}
\caption{\small \label{tab:systematics} Summary of the absolute systematic uncertainties on the measured observables.}
\begin{tabular}{@{}lllll@{}}
Source&
${\cal{A}}^{KK}_{d}$ & 
${\cal{A}}_d^{\rm fav}$ & 
${\cal{A}}^{KK}_{s}$ & 
${\cal{R}}^{KK}_{d}$ 
\\
\midrule

Production asymmetry & 
0.005 & 
0.006 & 
$-$ & 
0.003 \\

PID efficiency & 
0.004 & 
0.008 & 
0.005 & 
0.014  
\\

Trigger efficiency & 
0.004 & 
0.001 &
0.005 & 
0.022 
 \\

Selection efficiency & 
$-$ & 
$-$ & 
$-$ & 
0.040 
\\

Bias correction &
0.004 &
$-$ &
0.001 &
0.013 \\

Low-mass background & 
0.017 & 
0.001 & 
0.004 & 
0.042
\\

$\Bz \to \D \rho^0$ cross-feed & 
0.001 & 
$-$ & 
0.002 & 
0.008 
\\

Signal description & 
0.001 & 
0.001 & 
0.001 & 
0.005 
\\

\D branching fractions &
$-$ &
$-$ &
$-$ &
0.022 \\

\midrule
Total & 
0.019 & 
0.010 & 
0.008 & 
0.069 
\\

\end{tabular}
\end{center}
\end{table}
\renewcommand{\arraystretch}{1.}

\section{Results and summary}

This paper reports the analysis of $\Bz \to \D\Kstarz$ decays using 1.0 \invfb of $pp$ collision data. 
Potential contributions to the decay amplitudes from the non-resonant $\Bz \to \D\Kp\pim$ mode are reduced by requiring that 
the \Kstarz reconstructed mass is within $\pm 50$\mevcc of the nominal mass and the absolute value of the cosine 
of the \Kstarz helicity angle is greater than 0.4.
The results for the \CP-violating observables are
\begin{alignat}{5}
& {\cal A}_d^{KK}       &\,=&\,& -0.45                      &&\, \pm \, 0.23 & \ ({\rm stat}) \pm 0.02\ ({\rm syst}), \notag \\
& {\cal A}_d^{\rm fav} &\,=&\,& -0.08                      && \, \pm \, 0.08  &\ ({\rm stat}) \pm 0.01\ ({\rm syst}), \notag \\
& {\cal A}_s^{KK}        &\,=&\,& 0.04                       &&\, \pm \, 0.16  &\ ({\rm stat}) \pm 0.01\ ({\rm syst}), \notag \\
& {\cal R}_d^{KK}       &\,=&\,& 1.36                       &&_{-\phantom{1}0.32\phantom{1}}^{+\phantom{1}0.37}  &\ ({\rm stat}) \pm 0.07\ ({\rm syst}). \notag
\end{alignat}
The value of ${\cal R}_d^{KK}$ takes 
into account the ratio of the branching fractions of $\Dz \to \Kp\Km$ to $\Dz\to \Km\pip$ decays~\cite{Nakamura:2010zzi}. 
The correlation between ${\cal A}_d^{KK} $ and ${\cal R}_d^{KK}$ is equal to 0.16 and the 
correlations between the other observables are negligible.

These are the first measurements of \CP asymmetries in 
\Bz and \Bsb to $\D \Kstarz$ decays with the neutral \D meson decaying into a \CP-even 
final state. 
Triggering, reconstructing and selecting a pure sample of these fully hadronic $B$ decays is challenging in a high rate and high track-multiplicity environment, especially in the forward direction of LHCb.
The present statistical limitations are due to a combination of several factors, the most important one being the trigger.
In order to keep the output rate below its maximum of 1~MHz, the current hardware trigger imposes relatively restrictive criteria on the minimum transverse momentum of hadrons, which affect the efficiency for fully-hadronic modes. 
This limitation is overcome in the proposed LHCb upgrade~\cite{CERN-LHCC-2011-001, *CERN-LHCC-2012-007} by reading out the detector at the maximum LHC bunch-crossing frequency of 40 MHz. 
With more data,
improved measurements of these and other quantities in $\Bz\to \D \Kstarz$ 
decays will result in important constraints on the angle $\gamma$ of the Unitarity Triangle.








\section*{Acknowledgements}

\noindent We express our gratitude to our colleagues in the CERN
accelerator departments for the excellent performance of the LHC. We
thank the technical and administrative staff at the LHCb
institutes. We acknowledge support from CERN and from the national
agencies: CAPES, CNPq, FAPERJ and FINEP (Brazil); NSFC (China);
CNRS/IN2P3 and Region Auvergne (France); BMBF, DFG, HGF and MPG
(Germany); SFI (Ireland); INFN (Italy); FOM and NWO (The Netherlands);
SCSR (Poland); ANCS/IFA (Romania); MinES, Rosatom, RFBR and NRC
``Kurchatov Institute'' (Russia); MinECo, XuntaGal and GENCAT (Spain);
SNSF and SER (Switzerland); NAS Ukraine (Ukraine); STFC (United
Kingdom); NSF (USA). We also acknowledge the support received from the
ERC under FP7. The Tier1 computing centres are supported by IN2P3
(France), KIT and BMBF (Germany), INFN (Italy), NWO and SURF (The
Netherlands), PIC (Spain), GridPP (United Kingdom). We are thankful
for the computing resources put at our disposal by Yandex LLC
(Russia), as well as to the communities behind the multiple open
source software packages that we depend on.



\addcontentsline{toc}{section}{References}
\bibliographystyle{LHCb}
\bibliography{main}

\ifx\mcitethebibliography\mciteundefinedmacro
\PackageError{LHCb.bst}{mciteplus.sty has not been loaded}
{This bibstyle requires the use of the mciteplus package.}\fi
\providecommand{\href}[2]{#2}
\begin{mcitethebibliography}{10}
\mciteSetBstSublistMode{n}
\mciteSetBstMaxWidthForm{subitem}{\alph{mcitesubitemcount})}
\mciteSetBstSublistLabelBeginEnd{\mcitemaxwidthsubitemform\space}
{\relax}{\relax}

\bibitem{bib:GL}
M.~Gronau and D.~London, \ifthenelse{\boolean{articletitles}}{{\it {How to
  determine all the angles of the unitarity triangle from $B_d^0 \to D K_{\rm
  S}$ and $B_s^0 \to D\phi$}},
  }{}\href{http://dx.doi.org/10.1016/0370-2693(91)91756-L}{Phys.\ Lett.\  {\bf
  B253} (1991) 483}\relax
\mciteBstWouldAddEndPuncttrue
\mciteSetBstMidEndSepPunct{\mcitedefaultmidpunct}
{\mcitedefaultendpunct}{\mcitedefaultseppunct}\relax
\EndOfBibitem
\bibitem{bib:GW}
M.~Gronau and D.~Wyler, \ifthenelse{\boolean{articletitles}}{{\it {On
  determining a weak phase from charged $B$ decay asymmetries}},
  }{}\href{http://dx.doi.org/10.1016/0370-2693(91)90034-N}{Phys.\ Lett.\  {\bf
  B265} (1991) 172}\relax
\mciteBstWouldAddEndPuncttrue
\mciteSetBstMidEndSepPunct{\mcitedefaultmidpunct}
{\mcitedefaultendpunct}{\mcitedefaultseppunct}\relax
\EndOfBibitem
\bibitem{Dunietz:1991yd}
I.~Dunietz, \ifthenelse{\boolean{articletitles}}{{\it {CP violation with
  self-tagging $B_d$ modes}},
  }{}\href{http://dx.doi.org/10.1016/0370-2693(91)91542-4}{Phys.\ Lett.\  {\bf
  B270} (1991) 75}\relax
\mciteBstWouldAddEndPuncttrue
\mciteSetBstMidEndSepPunct{\mcitedefaultmidpunct}
{\mcitedefaultendpunct}{\mcitedefaultseppunct}\relax
\EndOfBibitem
\bibitem{Gronau:2002mu}
M.~Gronau, \ifthenelse{\boolean{articletitles}}{{\it {Improving bounds on
  $\gamma$ in $B^\pm \to DK^\pm$ and $B^{\pm,0} \to DX_{s}^{\pm,0}$}},
  }{}\href{http://dx.doi.org/10.1016/S0370-2693(03)00192-8}{Phys.\ Lett.\  {\bf
  B557} (2003) 198}, \href{http://arxiv.org/abs/hep-ph/0211282}{{\tt
  arXiv:hep-ph/0211282}}\relax
\mciteBstWouldAddEndPuncttrue
\mciteSetBstMidEndSepPunct{\mcitedefaultmidpunct}
{\mcitedefaultendpunct}{\mcitedefaultseppunct}\relax
\EndOfBibitem
\bibitem{Adeva:2009ny}
LHCb collaboration, B.~Adeva {\em et~al.},
  \ifthenelse{\boolean{articletitles}}{{\it {Roadmap for selected key
  measurements of LHCb}}, }{}\href{http://arxiv.org/abs/0912.4179}{{\tt
  arXiv:0912.4179}}\relax
\mciteBstWouldAddEndPuncttrue
\mciteSetBstMidEndSepPunct{\mcitedefaultmidpunct}
{\mcitedefaultendpunct}{\mcitedefaultseppunct}\relax
\EndOfBibitem
\bibitem{Aaij:2012kz}
LHCb Collaboration, R.~Aaij {\em et~al.},
  \ifthenelse{\boolean{articletitles}}{{\it {Observation of CP violation in
  $B^+$ to $DK^+$ decays}},
  }{}\href{http://dx.doi.org/10.1016/j.physletb.2012.04.060}{Phys.\ Lett.\
  {\bf B712} (2012) 203}, \href{http://arxiv.org/abs/1203.3662}{{\tt
  arXiv:1203.3662}}, erratum
  \href{http://dx.doi.org/10.1016/j.physletb.2012.05.060}{Phys. Lett.
  \textbf{B713} (2012) 351}\relax
\mciteBstWouldAddEndPuncttrue
\mciteSetBstMidEndSepPunct{\mcitedefaultmidpunct}
{\mcitedefaultendpunct}{\mcitedefaultseppunct}\relax
\EndOfBibitem
\bibitem{Aaij:1483187}
R.~Aaij {\em et~al.}, \ifthenelse{\boolean{articletitles}}{{\it {A
  model-independent Dalitz plot analysis of $B^\pm \to D K^\pm$ with $D \to
  K^0_{\rm S} h^+h^-$ ($h=\pi, K$) decays and constraints on the CKM angle
  $\gamma$}}, }{}\href{http://dx.doi.org/10.1016/j.physletb.2012.10.020}{Phys.\
  Lett.\  {\bf B718} (2012) 43}, \href{http://arxiv.org/abs/1209.5869}{{\tt
  arXiv:1209.5869}}\relax
\mciteBstWouldAddEndPuncttrue
\mciteSetBstMidEndSepPunct{\mcitedefaultmidpunct}
{\mcitedefaultendpunct}{\mcitedefaultseppunct}\relax
\EndOfBibitem
\bibitem{LHCb-PAPER-2011-008}
LHCb collaboration, R.~Aaij {\em et~al.},
  \ifthenelse{\boolean{articletitles}}{{\it {First observation of the decay
  $\Bsb \to \Dz \Kstarz$ and a measurement of the ratio of branching fractions
  $\frac{{\cal B}(\Bsb \to \Dz \Kstarz)}{{\cal B}(\Bzb \to \Dz \rho^0)}$}},
  }{}\href{http://dx.doi.org/10.1016/j.physletb.2011.10.073}{Phys.\ Lett.\
  {\bf B706} (2011) 32}, \href{http://arxiv.org/abs/1110.3676}{{\tt
  arXiv:1110.3676}}\relax
\mciteBstWouldAddEndPuncttrue
\mciteSetBstMidEndSepPunct{\mcitedefaultmidpunct}
{\mcitedefaultendpunct}{\mcitedefaultseppunct}\relax
\EndOfBibitem
\bibitem{Alves:2008zz}
LHCb collaboration, A.~A. Alves~Jr. {\em et~al.},
  \ifthenelse{\boolean{articletitles}}{{\it {The LHCb detector at the LHC}},
  }{}\href{http://dx.doi.org/10.1088/1748-0221/3/08/S08005}{JINST {\bf 3}
  (2008) S08005}\relax
\mciteBstWouldAddEndPuncttrue
\mciteSetBstMidEndSepPunct{\mcitedefaultmidpunct}
{\mcitedefaultendpunct}{\mcitedefaultseppunct}\relax
\EndOfBibitem
\bibitem{Aaij:2012me}
R.~Aaij {\em et~al.}, \ifthenelse{\boolean{articletitles}}{{\it {The \lhcb
  trigger and its performance}}, }{}\href{http://arxiv.org/abs/1211.3055}{{\tt
  arXiv:1211.3055}}\relax
\mciteBstWouldAddEndPuncttrue
\mciteSetBstMidEndSepPunct{\mcitedefaultmidpunct}
{\mcitedefaultendpunct}{\mcitedefaultseppunct}\relax
\EndOfBibitem
\bibitem{Aaij:1464076}
R.~Aaij {\em et~al.}, \ifthenelse{\boolean{articletitles}}{{\it {Observation of
  $\Bz \to \Dzb \Kp \Km$ and evidence of $\Bs \to \Dzb \Kp \Km$}},
  }{}\href{http://dx.doi.org/10.1103/PhysRevLett.109.131801}{Phys.\ Rev.\
  Lett.\  {\bf 109} (2012) 131801}, \href{http://arxiv.org/abs/1207.5991}{{\tt
  arXiv:1207.5991}}\relax
\mciteBstWouldAddEndPuncttrue
\mciteSetBstMidEndSepPunct{\mcitedefaultmidpunct}
{\mcitedefaultendpunct}{\mcitedefaultseppunct}\relax
\EndOfBibitem
\bibitem{Nakamura:2010zzi}
Particle Data Group, J.~Beringer {\em et~al.},
  \ifthenelse{\boolean{articletitles}}{{\it {Review of particle physics}},
  }{}\href{http://dx.doi.org/10.1103/PhysRevD.86.010001}{Phys.\ Rev.\  {\bf
  D86} (2012) 010001}\relax
\mciteBstWouldAddEndPuncttrue
\mciteSetBstMidEndSepPunct{\mcitedefaultmidpunct}
{\mcitedefaultendpunct}{\mcitedefaultseppunct}\relax
\EndOfBibitem
\bibitem{Sjostrand:2006za}
T.~Sj\"{o}strand, S.~Mrenna, and P.~Skands,
  \ifthenelse{\boolean{articletitles}}{{\it {PYTHIA 6.4 physics and manual}},
  }{}\href{http://dx.doi.org/10.1088/1126-6708/2006/05/026}{JHEP {\bf 05}
  (2006) 026}, \href{http://arxiv.org/abs/hep-ph/0603175}{{\tt
  arXiv:hep-ph/0603175}}\relax
\mciteBstWouldAddEndPuncttrue
\mciteSetBstMidEndSepPunct{\mcitedefaultmidpunct}
{\mcitedefaultendpunct}{\mcitedefaultseppunct}\relax
\EndOfBibitem
\bibitem{LHCb-PROC-2010-056}
I.~Belyaev {\em et~al.}, \ifthenelse{\boolean{articletitles}}{{\it {Handling of
  the generation of primary events in \gauss, the \lhcb simulation framework}},
  }{}\href{http://dx.doi.org/10.1109/NSSMIC.2010.5873949}{Nuclear Science
  Symposium Conference Record (NSS/MIC) {\bf IEEE} (2010) 1155}\relax
\mciteBstWouldAddEndPuncttrue
\mciteSetBstMidEndSepPunct{\mcitedefaultmidpunct}
{\mcitedefaultendpunct}{\mcitedefaultseppunct}\relax
\EndOfBibitem
\bibitem{Lange:2001uf}
D.~J. Lange, \ifthenelse{\boolean{articletitles}}{{\it {The EvtGen particle
  decay simulation package}},
  }{}\href{http://dx.doi.org/10.1016/S0168-9002(01)00089-4}{Nucl.\ Instrum.\
  Meth.\  {\bf A462} (2001) 152}\relax
\mciteBstWouldAddEndPuncttrue
\mciteSetBstMidEndSepPunct{\mcitedefaultmidpunct}
{\mcitedefaultendpunct}{\mcitedefaultseppunct}\relax
\EndOfBibitem
\bibitem{Allison:2006ve}
GEANT4 collaboration, J.~Allison {\em et~al.},
  \ifthenelse{\boolean{articletitles}}{{\it {Geant4 developments and
  applications}}, }{}\href{http://dx.doi.org/10.1109/TNS.2006.869826}{IEEE
  Trans.\ Nucl.\ Sci.\  {\bf 53} (2006) 270}\relax
\mciteBstWouldAddEndPuncttrue
\mciteSetBstMidEndSepPunct{\mcitedefaultmidpunct}
{\mcitedefaultendpunct}{\mcitedefaultseppunct}\relax
\EndOfBibitem
\bibitem{Agostinelli:2002hh}
GEANT4 collaboration, S.~Agostinelli {\em et~al.},
  \ifthenelse{\boolean{articletitles}}{{\it {GEANT4: a simulation toolkit}},
  }{}\href{http://dx.doi.org/10.1016/S0168-9002(03)01368-8}{Nucl.\ Instrum.\
  Meth.\  {\bf A506} (2003) 250}\relax
\mciteBstWouldAddEndPuncttrue
\mciteSetBstMidEndSepPunct{\mcitedefaultmidpunct}
{\mcitedefaultendpunct}{\mcitedefaultseppunct}\relax
\EndOfBibitem
\bibitem{LHCb-PROC-2011-006}
M.~Clemencic {\em et~al.}, \ifthenelse{\boolean{articletitles}}{{\it {The \lhcb
  simulation application, \gauss: design, evolution and experience}},
  }{}\href{http://dx.doi.org/10.1088/1742-6596/331/3/032023}{{J.\ of Phys:
  Conf.\ Ser.\ } {\bf 331} (2011) 032023}\relax
\mciteBstWouldAddEndPuncttrue
\mciteSetBstMidEndSepPunct{\mcitedefaultmidpunct}
{\mcitedefaultendpunct}{\mcitedefaultseppunct}\relax
\EndOfBibitem
\bibitem{Cranmer:2000du}
K.~S. Cranmer, \ifthenelse{\boolean{articletitles}}{{\it {Kernel estimation in
  high-energy physics}},
  }{}\href{http://dx.doi.org/10.1016/S0010-4655(00)00243-5}{Comput.\ Phys.\
  Commun.\  {\bf 136} (2001) 198},
  \href{http://arxiv.org/abs/hep-ex/0011057}{{\tt arXiv:hep-ex/0011057}}\relax
\mciteBstWouldAddEndPuncttrue
\mciteSetBstMidEndSepPunct{\mcitedefaultmidpunct}
{\mcitedefaultendpunct}{\mcitedefaultseppunct}\relax
\EndOfBibitem
\bibitem{LHCb-PAPER-2011-029}
LHCb collaboration, R.~Aaij {\em et~al.},
  \ifthenelse{\boolean{articletitles}}{{\it {First evidence of direct CP
  violation in charmless two-body decays of $B_s$ mesons}},
  }{}\href{http://dx.doi.org/10.1103/PhysRevLett.108.201601}{Phys.\ Rev.\
  Lett.\  {\bf 108} (2012) 201601}, \href{http://arxiv.org/abs/1202.6251}{{\tt
  arXiv:1202.6251}}\relax
\mciteBstWouldAddEndPuncttrue
\mciteSetBstMidEndSepPunct{\mcitedefaultmidpunct}
{\mcitedefaultendpunct}{\mcitedefaultseppunct}\relax
\EndOfBibitem
\bibitem{CERN-LHCC-2011-001}
LHCb collaboration, R.~Aaij {\em et~al.},
  \ifthenelse{\boolean{articletitles}}{{\it {Letter of intent for the LHCb
  upgrade}}, }{}
  \href{http://cdsweb.cern.ch/search?p=LHCb-PUB-2011-001&f=reportnumber&action%
_search=Search&c=LHCb+Reports&c=LHCb+Conference+Proceedings&c=LHCb+Conference+%
Contributions&c=LHCb+Notes&c=LHCb+Theses&c=LHCb+Papers}
  {LHCb-PUB-2011-001}\relax
\mciteBstWouldAddEndPuncttrue
\mciteSetBstMidEndSepPunct{\mcitedefaultmidpunct}
{\mcitedefaultendpunct}{\mcitedefaultseppunct}\relax
\EndOfBibitem
\bibitem{CERN-LHCC-2012-007}
LHCb collaboration, I.~Bediaga {\em et~al.},
  \ifthenelse{\boolean{articletitles}}{{\it {Framework TDR for the LHCb
  upgrade}}, }{}
  \href{http://cdsweb.cern.ch/search?p=CERN-LHCC-2012-007&f=reportnumber&actio%
n_search=Search&c=LHCb+Reports&c=LHCb+Conference+Proceedings&c=LHCb+Conference%
+Contributions&c=LHCb+Notes&c=LHCb+Theses&c=LHCb+Papers}
  {CERN-LHCC-2012-007}\relax
\mciteBstWouldAddEndPuncttrue
\mciteSetBstMidEndSepPunct{\mcitedefaultmidpunct}
{\mcitedefaultendpunct}{\mcitedefaultseppunct}\relax
\EndOfBibitem
\end{mcitethebibliography}

\end{document}